\documentclass[%
 aps,pre,
 amsmath,amssymb,
 reprint,%
]{revtex4-2}

\usepackage{graphicx}% Include figure files
\usepackage{dcolumn}% Align table columns on decimal point
\usepackage{bm}% bold math
\usepackage{amsfonts,amsthm,mathbbol}
\usepackage[utf8]{inputenc}
\usepackage[T1]{fontenc}
\usepackage{mathptmx}

\begin{document}

\title[Infinite ergodicity for geometric Brownian motion]{Infinite ergodicity for geometric Brownian motion}

\author{Stefano Giordano}%
\email{stefano.giordano@univ-lille.fr}  
\affiliation{University of Lille, CNRS, Centrale Lille, Univ. Polytechnique Hauts-de-France, UMR 8520 - IEMN - Institut d'{\'E}lectronique, de Micro{\'e}lectronique et de Nanotechnologie, F-59000 Lille, France
}

\author{Fabrizio Cleri}%
\email{fabrizio.cleri@univ-lille.fr}  
\affiliation{University of Lille, Institut d'{\'E}lectronique, de Micro{\'e}lectronique et de Nanotechnologie (IEMN CNRS UMR8520) and Departement de Physique, F-59652 Villeneuve d'Ascq, France
}

 \author{Ralf Blossey}
\email{ralf.blossey@univ-lille.fr}
\affiliation{University of Lille, Unit{\'e} de Glycobiologie Structurale et Fonctionnelle (UGSF), CNRS UMR8576, F-59000 Lille, France}

\date{\today}% It is always \today, today,

%%%%%%%%%%%%%%%%%%%%%%%%%%%%%%%%%%%%%%%%%%%%%%%%%%%%%%%%%%%%%%%%%%%%%%%%%%%%%%%%
\begin{abstract}
Geometric Brownian motion is an exemplary stochastic processes obeying multiplicative noise, with widespread applications in several fields, e.g. in finance, in 
physics and biology. The definition of the process depends crucially on the interpretation of the stochastic integrals which involves 
the discretization parameter $\alpha$ with $0 \leq \alpha \leq 1$ , giving rise to the well-known special cases $\alpha=0$ (It\^{o}), $\alpha=1/2$ (Fisk-Stratonovich) 
and $\alpha=1$ (H\"{a}nggi-Klimontovich or anti-It\^{o}). In this paper we study the asymptotic limits of the probability distribution functions (PDFs) of geometric Brownian 
motion and some related generalizations. We establish the conditions for the existence of normalizable asymptotic distributions depending on the discretization 
parameter $\alpha$. Using the infinite ergodicity approach, recently applied to stochastic processes with multiplicative noise by E. Barkai and collaborators, we show 
how meaningful asymptotic results can be formulated in a transparent way.
\end{abstract}

\maketitle

%%%%%%%%%%%%%%%%%%%%%%%%%%%%%%%%%%%%%%%%%%%%%%%%%%%%%%%%%%%%%%%%%%%%%%%%%%%%%%%%
\section{Introduction}

Stochastic processes in the presence of multiplicative noise are a commonly encountered phenomenon in the sciences. 
In a general way, we consider a variable $x(t)$ which follows a stochastic differential equation. If the corresponding random term in the equation does not depend on the state of the system (i.e., 
on $x(t)$), we call it additive noise. On the other hand, if the random term depends on the state of the system $x(t)$, then the noise term is called multiplicative.
A prime example of multiplicative noise from physics is the statistical theory of turbulence, where the energy cascade can be modeled by a Markov process at least down to the Taylor scale, which governs the intermediate scale of turbulent eddies \cite{fuchs2022}. 
The one-dimensional stochastic process across the scales, the Kolmogoroff-Obukhov theory (K62) \cite{birnir2013} can be approximated by a geometric Brownian process. The geometric Brownian process is defined by a stochastic differential equation with the random term which is directly proportional to the state $x(t)$ of the system.
A highly prominent example of geometric Brownian motion
is the modeling of stock prices, notably with the celebrated Black-Scholes model of option pricing \cite{hull2021,bouchaud2009}. The relation 
between the stochastic behaviour of financial markets and the turbulence cascade has been discussed in Ref.\cite{mantegna2000}. 
Multiplicative noises are also used for explaining the ballistic-to-diffusive transition of the heat propagation \cite{landi2014,palla2020}. In this case, a chain of particles is considered with a multiplicative stochastic force field able to be energy-conserving for each particle of the system.
Random systems with multiplicative noise find applications also to cosmology and statistical field theory \cite{habib1993}.
Finally, there are several applications in biology for which we cite as examples the stochastic firing of neurons \cite{bauermann2019,zhu}, phenotypic variability and
gene expression \cite{gen1,gen2}, and the motion of molecular motors \cite{baule2008}, notably chromatin remodeling complexes acting on nucleosomes \cite{blossey2019,breoni2021}.  

All these cases have in common that they deal with first-order stochastic differential equations of the form
\begin{equation}
\frac{\mathrm{d}x}{\mathrm{d}t}=h(x,t)+g(x,t)\xi(t),
\end{equation}
where $h(x,t)$ is the drift term, $g(x,t)$ is the diffusion term, and the stochastic process $\xi(t)$ is a Gaussian noise with average value $E\left\lbrace \xi(t)\right\rbrace=0 $, and correlation $E\left\lbrace \xi(t)\xi(\tau)\right\rbrace=2\delta(t-\tau) $.
The stochastic differential equation has a well-defined meaning only if we declare the adopted interpretation
of the stochastic integrals. To achieve this, we have to specify the discretization parameter $\alpha$, defining the position of the point at which we 
calculate any integrated function in the small intervals of the adopted Riemann sum ($0 \leq \alpha \leq 1$) \cite{coffey1,risken,oksendal2003,coffey2}. 
This integration theory includes the It\^{o} ($\alpha=0$) \cite{ito}, the Fisk-Stratonovich ($\alpha=1/2$) \cite{fisk,strato}, and the H\"{a}nggi-Klimontovich or anti-It\^{o} ($\alpha=1$) \cite{thomas,klimo} 
as particular cases (see also Ref.\cite{soko}). 
In fact, our above-mentioned example of the turbulent cascade, in which $t$ is identified with the cascade scale,
is commonly interpreted in the Fisk-Stratonovich sense, while the Black-Scholes stock market model is treated in the It{\^o}-interpretation. For the heat conduction, it has been proved that all stochastic interpretations are equivalent \cite{palla2020}.

The stochastic process can likewise be described with the help of the Fokker-Planck equation, a partial differential equation for the probability density function (PDF) $W(x,t)$ 
of the stochastic process given by \cite{coffey1,risken,oksendal2003,coffey2,denisov2009}
\begin{equation}
\frac{\partial W}{\partial t}=-\frac{\partial }{\partial x}\left[ \left(h+2\alpha g \frac{\partial g}{\partial x}\right) \right] +\frac{\partial^2}{\partial x^2}\left(g^2 W \right) ,
\label{FP1}
\end{equation}
where the first term represents the force-dependent drift, the second a noise-induced drift which explicitly depends on $\alpha$, and the third the diffusion term
generated by the noise. The noise-induced drift term is absent when $\partial g/\partial x = 0$, i.e. for purely additive noise. Thus, 
the choice of the stochastic calculus is relevant only in the case of multiplicative noise. The theory can be generalized to take into consideration possible 
cross-correlation of the noises \cite{deni1,deni2}. The Fokker-Planck Eq.(\ref{FP1}) can be also rewritten in the following useful form
\begin{equation}
\frac{\partial W}{\partial t}=\frac{\partial }{\partial x}\left\lbrace  -hW+ g^{2\alpha}\frac{\partial }{\partial x}\left[ g^{2(1-\alpha)}W\right] \right\rbrace,
\label{FP2}
\end{equation}
which is readily demonstrated by performing the derivatives.

In this paper we are interested in a full characterization of the PDF for geometric Brownian motion and some generalizations of this process.
We will in particular consider the class of stochastic equations with simple algebraic nonlinearities for the drift and noise terms
\begin{equation}
\frac{\mathrm{d}x}{\mathrm{d}t}= H(t) x^n + G(t) x^m \xi(t),
\label{alg}
\end{equation}
since they will readily allow us to obtain analytic results. Specifically, we are interested in the conditions that guarantee the existence of a normalizable asymptotic long-time limit, 
or stationary PDFs, given by
\begin{equation}
W_{as}(x) = \lim_{t \rightarrow \infty} W(x,t)\, .
\label{as}
\end{equation}
That the existence of such PDFs is not generally guaranteed, and it indeed depends on the discretization parameter $\alpha$, was recently shown by Barkai 
and collaborators \cite{bar1,bar2,bar3} for certain cases we comment on below. The authors introduced the concept of infinite ergodicity in the discussion, which allowed 
them to define a procedure to extract meaningful physical quantities from these non-normalizable distributions. 
In particular, it is possible to determine the asymptotic behavior (with time going to infinity) of the expected value of different physical observables.  In statistical mechanics, these approaches are used when the potential energy is non-confining and thus generates an infinite phase space (or infinite measure space \cite{aaronson}), whence the name infinite ergodicity. 
In Ref.\cite{bar2}, the case of geometric Brownian motion 
was explicitly excluded from the discussion, so we extend its analysis in the present work. More specifically, we consider a geometric Brownian motion, $m=1$ in Eq.(\ref{alg}), with a nonlinear drift ($n\neq 1$) and then we introduce a generalization with a nonlinear diffusion term ($m\neq 1$).

The structure of the paper follows. In Section \ref{sectvgbm}, we introduce the geometric Brownian motion with time-varying and linear drift and diffusion terms. We obtain here a generalized log-normal distribution. In Section \ref{nonlineardrift}, we introduce a nonlinear drift term in the geometric Brownian  motion stochastic equation, and we investigate the existence of normalizable asymptotic densities as defined in Eq.(\ref{as}). In Section \ref{iet}, we discuss the concept of infinite ergodicity by considering a simple overdamped system taken from statistical mechanics. We then apply this concept to the geometric Brownian motion with non linear drift term in Section \ref{ietgbm}. To conclude, we generalize our approach for systems with an algebraic nonlinear diffusion term in Section \ref{further}.

\section{Time-varying geometric Brownian motion} 
\label{sectvgbm}

We initially focus on geometric Brownian motion where the functions $h(x,t)$ and $g(x,t)$ are proportional to $x$ \cite{saaty,gardiner,vik}.
Thus we consider the time-varying geometric Brownian motion characterized by the stochastic Eq.(\ref{alg}) with $ n = m =1$,
\begin{equation}
\frac{\mathrm{d}x}{\mathrm{d}t}=H(t)x+G(t)x\xi(t),
\label{tvgbm}
\end{equation}
where $H(t)$ and $G(t)$ are two arbitrary time-dependent functions. The Fokker-Planck Eqs.(\ref{FP1}) and (\ref{FP2}) can be written in this case as
\begin{equation}
\frac{\partial W}{\partial t}=-\left(H+2\alpha G^2\right) \frac{\partial }{\partial x}\left(xW \right) +G^2\frac{\partial^2}{\partial x^2}\left(x^2 W \right) ,
\label{FP1gbm}
\end{equation}
and
\begin{equation}
\frac{\partial W}{\partial t}=-H\frac{\partial }{\partial x}\left(xW \right)  + G^2\frac{\partial }{\partial x}\left\lbrace x^{2\alpha}\frac{\partial }{\partial x}\left[ x^{2(1-\alpha)}W\right]\right\rbrace.
\label{FP2gbm}
\end{equation}
We are interested in finding the general solution of these equations for arbitrary functions $H(t)$ and $G(t)$. 
The driftless case $H(t)=0$ and with constant $G(t) \equiv G_0 $ is described by a log-normal distribution \cite{saaty}
\begin{equation}
f_\mathbf{x}(x)=\frac{1}{x\sqrt{2\pi \sigma^2}}e^{-\frac{\left(\log x-\mu \right)^2 }{2\sigma^2}},
\label{lnd}
\end{equation}
defined on the positive real line $ x > 0$, with suitable real parameters $\sigma$ and $\mu$ that define the shape of the distribution.  
The first-order and second-order expectation values are given by the expressions
\begin{eqnarray}
\label{exlnd}
E\left\lbrace \mathbf{x}\right\rbrace = e^{\mu+\frac{\sigma^2}{2}}\, ,\,\, 
%\label{ex2lnd}
E\left\lbrace \mathbf{x}^2\right\rbrace = e^{2\mu+2\sigma^2},
\end{eqnarray}and the variance is given by  
\begin{equation}
\sigma_\mathbf{x}^2=E\left\lbrace \mathbf{x}^2\right\rbrace- E\left\lbrace \mathbf{x}\right\rbrace ^2=\left( e^{\sigma^2}-1\right) e^{2\mu+\sigma^2}.
\end{equation}
From Eqs.(\ref{exlnd}), we deduce the parameters $\mu$ and $\sigma^2$ as function of the expectation values as 
\begin{eqnarray}
\label{mu}
\mu = \log\frac{E\left\lbrace \mathbf{x}^2\right\rbrace}{E\left\lbrace \mathbf{x}\right\rbrace ^2}\, ,\,\,
%\label{si}
\sigma^2 = \log\frac{E\left\lbrace \mathbf{x}\right\rbrace ^2}{\sqrt {E\left\lbrace \mathbf{x}^2\right\rbrace}}.
\end{eqnarray}
We now assume that the solution of the Fokker-Planck equation has a log-normal form also for arbitrary functions $H(t)$ and $G(t)$.
This leads to the following evolution equations for the expectation values: 
\begin{eqnarray}
\label{av1}
\frac{dE\left\lbrace \mathbf{x}\right\rbrace}{dt}&=&\left[H(t)+2\alpha G^2(t) \right] E\left\lbrace \mathbf{x}\right\rbrace,\\
\label{av2}
\frac{dE\left\lbrace \mathbf{x}^2\right\rbrace}{dt}&=&2\left[H(t)+(2\alpha+1) G^2(t) \right] E\left\lbrace \mathbf{x}^2\right\rbrace.
\end{eqnarray}
These equations were obtained by multiplying the Fokker-Planck equation by $x$ and by $x^2$ and integrating the results on the interval $(0, \infty)$.
An integration by parts eventually leads to Eqs.(\ref{av1}) and (\ref{av2}). These differential equations can be solved to obtain
\begin{eqnarray}
\label{exp1}
E\left\lbrace \mathbf{x}\right\rbrace&=&\mu_0e^{\int_0^t\left[H(u)+2\alpha G^2(u) \right]\mathrm{d}u} ,\\
\label{exp2}
E\left\lbrace \mathbf{x}^2\right\rbrace&=&\left(\mu_0^2+\sigma_0^2 \right) e^{2\int_0^t\left[H(u)+(2\alpha+1) G^2(u) \right]\mathrm{d}u},
\end{eqnarray}
where $\mu_0$ and $\sigma_0^2$ are the average value and the variance of $\mathbf{x}$ for $t=0$, respectively.
Substituting Eqs.(\ref{exp1}) and (\ref{exp2}) into Eq.(\ref{mu}), we get
\begin{eqnarray}
\nonumber
\mu&=&\frac{1}{2}\log\frac{\mu_0^4}{\mu_0^2+\sigma_0^2}+\int_0^t\left[H(u)+(2\alpha -1) G^2(u) \right]\mathrm{d}u,\\\label{muf}\\
\label{sif}
\sigma^2 &=&\log\frac{\mu_0^2+\sigma_0^2}{\mu_0^2}+2\int_0^t G^2(u) \mathrm{d}u.
\end{eqnarray}
In particular, if $\sigma_0=0$, we have
\begin{eqnarray}
\label{muf0}
\mu&=&\log\mu_0+\int_0^t\left[H(u)+(2\alpha -1) G^2(u) \right]\mathrm{d}u,\\
\label{sif0}
\sigma^2 &=&2\int_0^t G^2(u) \mathrm{d}u.
\end{eqnarray} 
In order to demonstrate that the corresponding log-normal distribution really is the exact solution of our problem,
the Fokker-Planck equation in Eq.(\ref{FP1gbm}) or Eq.(\ref{FP2gbm}), with the initial condition $W(x,0)=\delta(x-\mu_0)$,
must be solved by the trial density
\begin{equation}
W(x,t)=\frac{\exp\left\lbrace {-\frac{\left[\log \frac{x}{\mu_0}-\int_0^t\left[H(u)+(2\alpha -1) G^2(u) \right]\mathrm{d}u \right]^2 }{4\int_0^t G^2(u) 
\mathrm{d}u}}\right\rbrace }{2x\sqrt{\pi \int_0^t G^2(u) \mathrm{d}u}},
\label{solfp}
\end{equation}
which follows from Eq.(\ref{lnd}) combined with Eqs.(\ref{muf0}) and (\ref{sif0}). This can be verified by a lengthy but straightforward calculation.
Our result in Eq.(\ref{solfp}) therefore generalizes the log-normal solution to the time-varying case independent of the interpretation of the stochastic integration
rule ($0 \leq \alpha \leq 1$). We remark that the obtained solution automatically satisfies the reflecting boundary condition at $x = 0$.

\begin{figure}[t]
\centering
\includegraphics[scale=0.6]{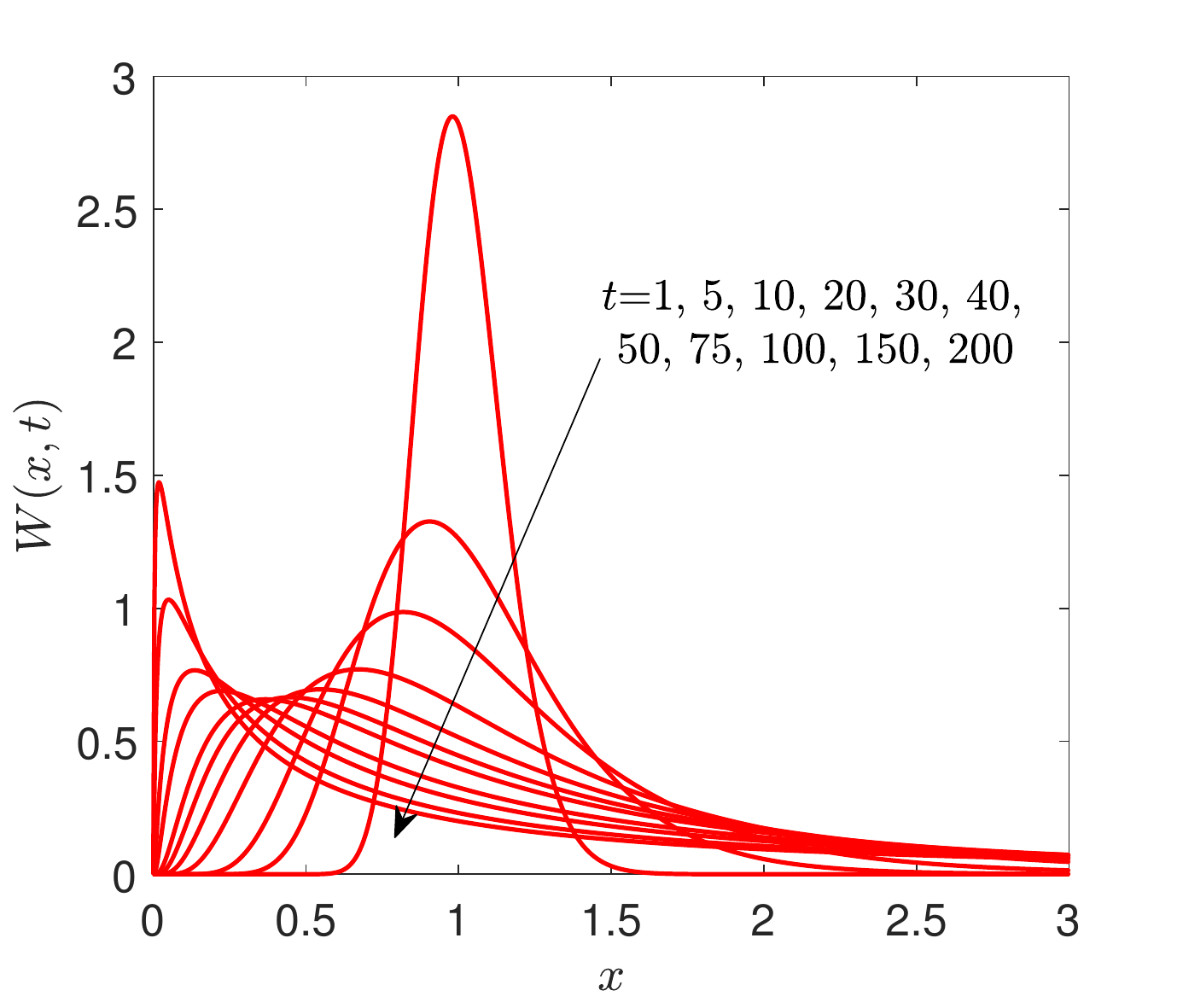}
\caption{\label{lognor} Example of log-normal distribution evolution. We implemented Eq.(\ref{solfp0}) with the parameters $H_0=0$, $G_0=1/10$, $\mu_0=1$, and $\alpha=1/2$.}
\end{figure}

While we have found a completely general expression for the PDF $W(x,t)$, a proper normalizable long-time limit for $t \rightarrow\infty $ of the PDF does
not always exist, depending on the form of $H(u)$ and $G(u)$. A normalizable stationary PDF exists only if the integrals $\int_0^t H(u)\mathrm{d}u$ and 
$\int_0^t G^2(u)\mathrm{d}u$ converge to finite values for $t\to\infty$. 
For example, if we take constant values for these functions, $H(u)=H_0 $ and $G(u)=G_0$, the 
density is given by
\begin{equation}
W(x,t)=\frac{\exp\left\lbrace {-\frac{\left[\log \frac{x}{\mu_0}-H_0t-(2\alpha -1) G_0^2 t \right]^2 }{4 G_0^2 t}}\right\rbrace }{2xG_0\sqrt{\pi t}},
\label{solfp0}
\end{equation}
which does not converge to an asymptotic or equilibrium distribution (see Fig.\ref{lognor}). This observation is the starting point for our following discussion. 

\section{Drift effect on geometric Brownian motion}
\label{nonlineardrift}

The result from the previous Section leads us to investigate whether a suitable nonlinear drift term can generate an asymptotic equilibrium density for a constant $G(t)=G_0 $. 
We thus now consider the stochastic differential equation
\begin{equation}
\frac{\mathrm{d}x}{\mathrm{d}t} = h(x) + G_0 x \xi(t),
\label{tvgbmdrift}
\end{equation}
where the drift term $h(x)$ is for the moment left unspecified. As before, we can associate the following Fokker-Planck equation, governing the evolution of the density $W(x,t)$
\begin{equation}
\frac{\partial W}{\partial t}=-\frac{\partial }{\partial x}\left(hW \right)  + G_0^2\frac{\partial }{\partial x}\left\lbrace x^{2\alpha}\frac{\partial }{\partial x}\left[ x^{2(1-\alpha)}W\right]\right\rbrace\, .
\label{FP2gbmdrift}
\end{equation}
The asymptotic solution of this Fokker-Planck equation, $W_{as}(x)=\lim_{t \to\infty}W(x,t)$, fulfills the equation
\begin{equation}
\label{asy}
0=-\left(hW_{as} \right)  + G_0^2 x^{2\alpha}\frac{\mathrm{d}}{\mathrm{d} x}\left[ x^{2(1-\alpha)}W_{as}\right]\, .
\end{equation}
Introducing $\Theta(x) = x^{2(1-\alpha)}W_{as}(x)$, we have the simpler equation
\begin{equation}
\frac{\mathrm{d}\Theta(x)}{\mathrm{d}x}=\frac{h(x)}{G_0^2x^2} \Theta(x)  ,
\label{eqtheta}
\end{equation}
whose solution is
\begin{equation}
\Theta(x)=\exp\left( \int\frac{h(x)}{G_0^2x^2} dx\right)\, .
\label{theta}
\end{equation}
Expressed in terms of the stationary PDF $W_{as}(x)$ we have
\begin{equation}
W_{as}(x)=\frac{K}{x^{2(1-\alpha)}}\exp\left( \int\frac{h(x)}{G_0^2x^2} dx\right),
\label{www}
\end{equation}
where we introduced a normalization constant $K$. 
Interestingly, this expression is reminiscent of the Pope-Ching result (for $\alpha=1/2$), stating a relationship between the PDF of any stationary process and the expectations of time derivatives
of the state of the system \cite{pope}. More precisely, this result indicates that both the expectations of the time
derivative squared and of the second time derivative define the PDF shape (see Eq.(7) in Ref.\cite{pope}). Interestingly, this result has been used to better understand turbulent flows data and has been thoroughly discussed in Ref.\cite{sokolov}.

The case of a linear drift term $ h(x) = H_0x$ leads back to the geometric Brownian motion already studied in Section \ref{sectvgbm}, 
for which we have already seen that there is no asymptotic solution. If $h(x) = H_0x$, Eq.(\ref{www}) immediately gives
\begin{equation}
W_{as}(x) = K x^{-2(1-\alpha)+H_0/G_0^2}\, ,
\label{www0}
\end{equation}
and it is readily seen that the integral $\int_0^\infty W_{as}(x)\mathrm{d}x$ cannot converge. 
In fact, we know that $\int_1^\infty x^{-k}\mathrm{d}x$ is convergent for $k>1$, and that $\int_0^1 x^{-k}\mathrm{d}x$ is convergent for $k<1$, therefore it is impossible to have values of  ${2(1-\alpha)-H_0/G_0^2}$ that render the integral $\int_0^\infty W_{as}(x)\mathrm{d}x$ convergent for  
$x\to 0$ and $x\to\infty$ at the same time.

It is therefore of interest to investigate whether different nonlinear forcing terms are able to generate a normalizable asymptotic density and consider
\begin{equation}
\label{drift}
h(x) = - H_0 x^n,
\end{equation}
where $n$ is a real number. The stochastic differential equation is now given by 
\begin{equation}
\frac{\mathrm{d}x}{\mathrm{d}t}= - H_0 x^n + G_0 x \xi(t)\, .
\label{tvgbmdriftspecial}
\end{equation}
In this case, Eq.(\ref{www}) yields
\begin{equation}
W_{as}(x)=\frac{K}{x^{2(1-\alpha)}}\exp\left(-\frac{H_0}{G_0^2}\frac{x^{n-1}}{n-1} \right),
\label{wwwa}
\end{equation}
where $n\neq 1$. In order to have a normalized asymptotic density, the inverse of the constant $K$ must be given by the integral
\begin{equation}
\frac{1}{K}=\int_0^{+\infty}\frac{1}{x^{2(1-\alpha)}}\exp\left(-\frac{H_0}{G_0^2}\frac{x^{n-1}}{n-1} \right)\mathrm{d}x,
\label{wwwak}
\end{equation}
the convergence of which must be carefully inspected. Since the integrand consists of the product of a algebraic and an exponential function, there 
arise two sets of conditions that can ensure convergence of the integral on the right hand side of Eq. (\ref{wwwak}): 
\\
(i) the term $x^{-2(1-\alpha)}$ is convergent for $x\to 0$ when $2(1-\alpha)<1$, or $\alpha>1/2$. In this case the exponential term must ensure the convergence for $x\to\infty$,
which is the case if $H_0>0$ and $n-1>0$. Indeed, in this case, $\exp\left(-\frac{H_0}{G_0^2}\frac{x^{n-1}}{n-1} \right)\to 0$ when $x\to\infty$. Finally, the integral in Eq.(\ref{wwwak}) 
is convergent if
\begin{equation}
\alpha>\frac{1}{2}, \,\,\,H_0>0, \mbox{ and } n-1>0;
\label{condnotgood}
\end{equation}
\\
(ii) The term $x^{-2(1-\alpha)}$ ensures the convergence for $x\to\infty$ if $2(1-\alpha) > 1$, or $\alpha < 1/2$. So, the exponential must handle the convergence for $x\to 0$. This is 
possible if $H_0 < 0$ and $n-1<0$. Indeed, in this case, $\exp\left(-\frac{H_0}{G_0^2}\frac{x^{n-1}}{n-1} \right)\to 0$ when $x\to 0$. Hence, the integral in Eq.(\ref{wwwak}) is convergent 
also if 
\begin{equation}
\alpha<\frac{1}{2}, \,\,\,H_0<0, \mbox{ and } n-1<0.
\label{condgood}
\end{equation}
\\
Within these two complementary regions of convergence we have the conditions that
\begin{equation}
G_0^2\frac{n-1}{H_0}>0, \mbox{ and } \frac{2\alpha-1}{n-1}>0\, .
\label{conds}
\end{equation}
An important finding is that we have found non-linear drift terms that are in fact able to generate an asymptotic equilibrium density even when $G(t) = G_0$.
It is interesting to notice that, however, for $\alpha=1/2$, the often invoked Fisk-Stratonovich case, the convergence condition cannot be fulfilled, and therefore we cannot use the 
Fisk Stratonovich interpretation of the stochastic calculus for Eq. (\ref{www0}) if the asymptotic density must remain normalisable. 

\begin{figure}[t]
\centering
\includegraphics[scale=0.6]{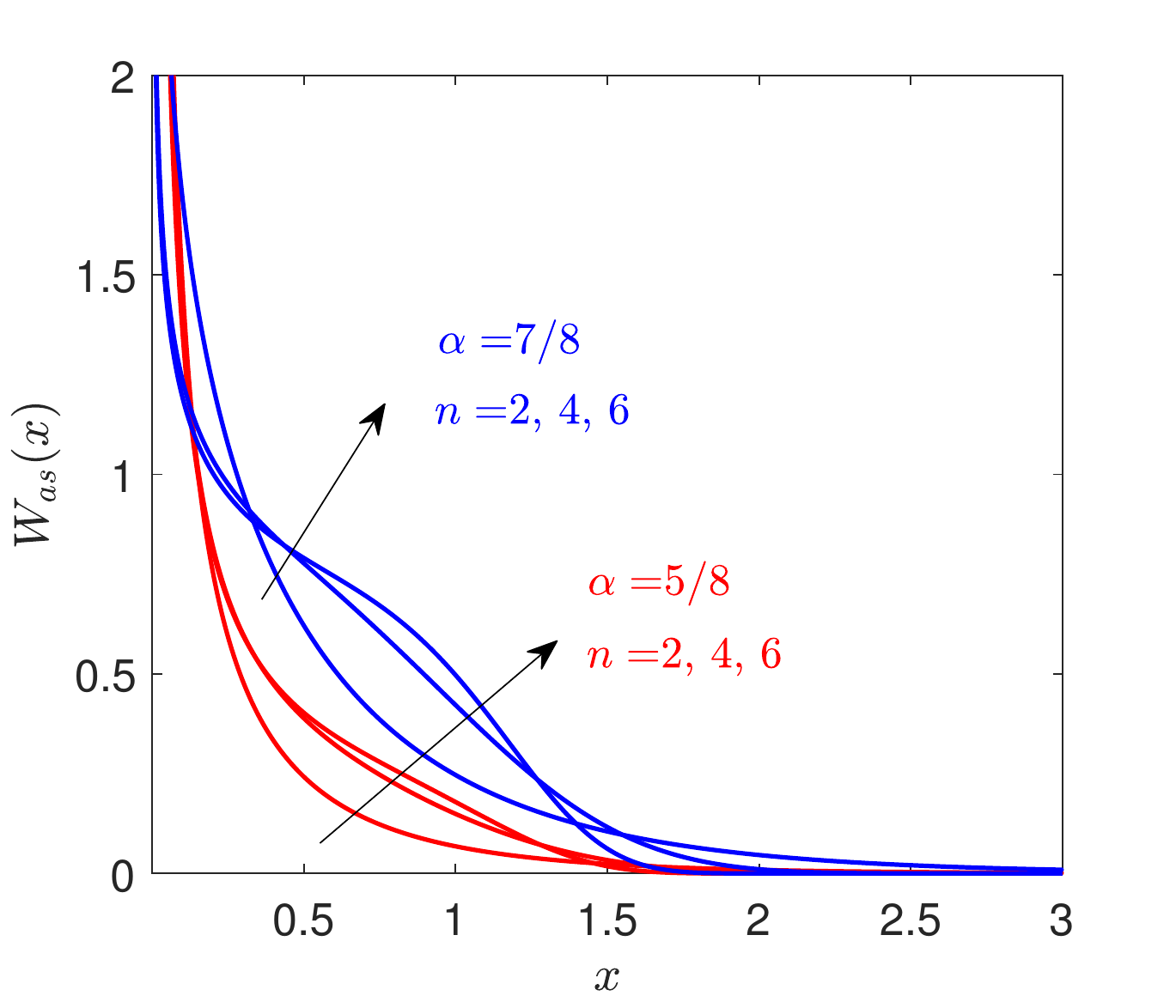}
\caption{\label{was2} Example of asymptotic distributions in the anti-It\^{o}-side region defined by Eq.(\ref{condnotgood}). We implemented Eq.(\ref{wwwafin}) with the parameters $H_0=3/2$ and $G_0=1$.}
\end{figure}

\begin{figure}[t]
\centering
\includegraphics[scale=0.6]{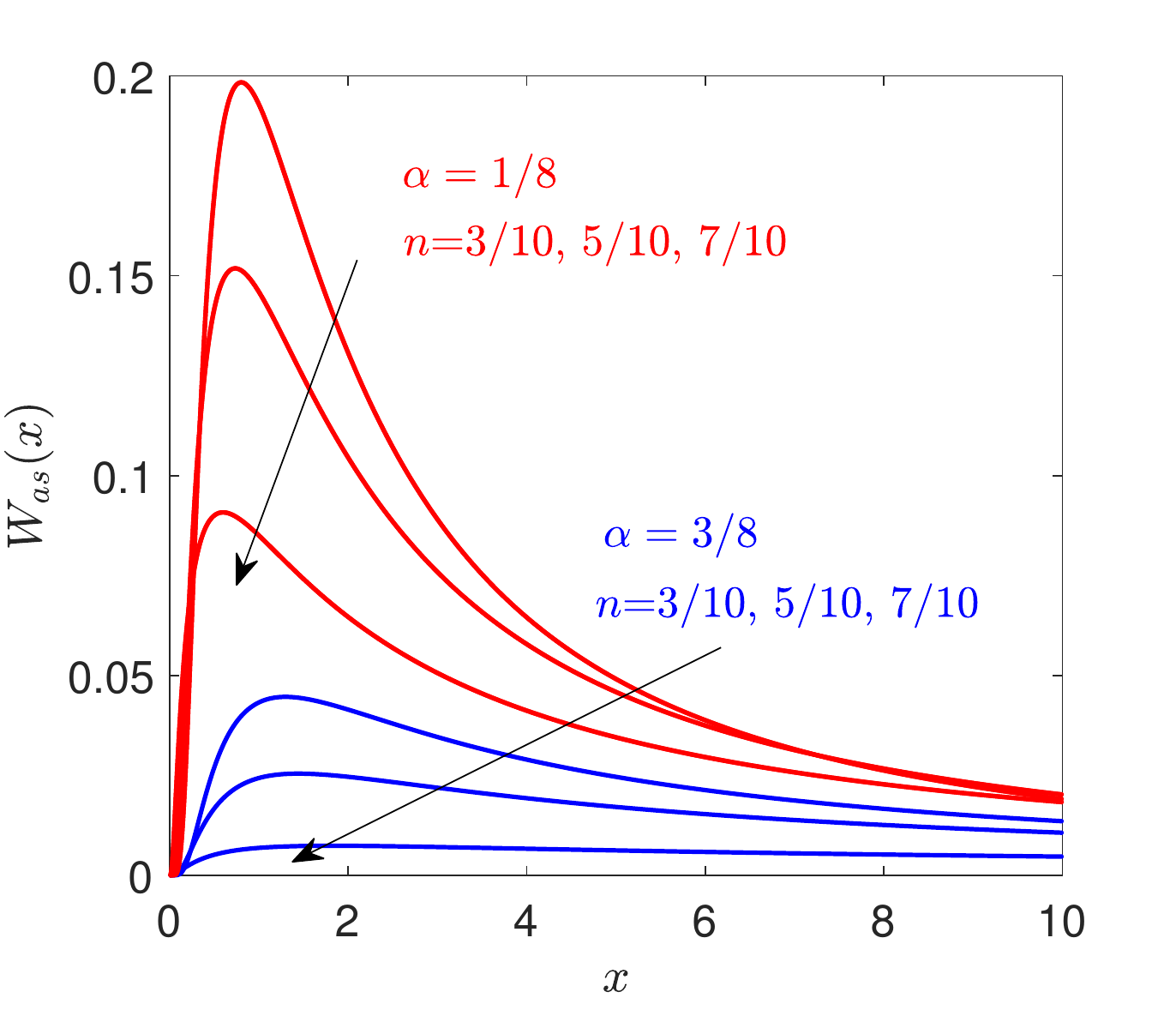}
\caption{\label{was1} Example of asymptotic distributions in the It\^{o}-side region defined by Eq.(\ref{condgood}). We implemented Eq.(\ref{wwwafin}) with the parameters $H_0=-3/2$ and $G_0=1$.}
\end{figure}

We will next try to obtain an explicit expression for the integral Eq. (\ref{wwwak}). To this aim, we introduce the substitution
\begin{equation}
t = \frac{H_0}{G_0^2}\frac{x^{n-1}}{n-1} \Leftrightarrow x=\left(\frac{n-1}{H_0}G_0^2 \right)^\frac{1}{n-1} t^\frac{1}{n-1},
\label{sub}
\end{equation}
which leads to
\begin{equation}
\frac{1}{K}=\frac{1}{\vert n-1 \vert}\left(\frac{n-1}{H_0} G_0^2\right)^\frac{2\alpha-1}{n-1}\int_0^{+\infty} t^{\frac{2\alpha-1}{n-1}-1}e^{-t}\mathrm{d}t.
\label{wwwakk}
\end{equation}
Here, we have included the term $\vert n-1 \vert$ for the following reason: if $n-1>0$, when $x\to 0$, we have $t\to 0$, and when $x\to\infty$, we have $t\to\infty$. 
Conversely, if $n-1<0$, when $x\to 0$, we have $t\to \infty$, and when $x\to\infty$, we have $t\to 0$. Hence, the order of integration changes depending on the 
sign of $n-1$. The integral in Eq. (\ref{wwwakk}) is of the form of the Gamma function, so that by using its definition \cite{math1,math2,math3}
 \begin{equation}
\Gamma(z)=\int_0^{+\infty} t^{z-1}e^{-t}\mathrm{d}t,
\end{equation}
we finally obtain from Eq.(\ref{wwwakk}) the result
\begin{equation}
\frac{1}{K}=\frac{1}{\vert n-1 \vert}\left(\frac{n-1}{H_0} G_0^2\right)^\frac{2\alpha-1}{n-1}\Gamma\left( \frac{2\alpha-1}{n-1}\right) .
\label{wwwakkk}
\end{equation}
Although we have shown that the integral is convergent in the anti-It\^{o}-side region defined by Eq.(\ref{condnotgood}), it is important to point out that for certain values of $n$ that are too high there may be problems with the existence of the solution of the stochastic equation and blow-up phenomena \cite{oksendal2003}. In any case, for the subsequent application of the infinite ergodicity concept, we will always use the It\^{o}-side region defined by conditions in Eq.(\ref{condgood}).

Anyway, the asymptotic density reads as 
\begin{equation}
W_{as}(x)=\frac{\vert n-1 \vert}{\left(\frac{n-1}{H_0} G_0^2\right)^\frac{2\alpha-1}{n-1}\Gamma\left( \frac{2\alpha-1}{n-1}\right)}\frac{\exp\left(-\frac{H_0}{G_0^2}\frac{x^{n-1}}{n-1} \right)}{x^{2(1-\alpha)}},
\label{wwwafin}
\end{equation}
which is valid when the conditions in Eq.(\ref{conds}) are fulfilled. 
In Figs.\ref{was2} and \ref{was1} we show the shape of the asymptotic density for different values of the parameters in the anti-It\^{o}-side and It\^{o}-side regions, defined by Eqs.(\ref{condnotgood}) and (\ref{condgood}), respectively. We note that the densities are singular for $x=0$ in the anti-It\^{o}-side region while they are regular everywhere for the It\^{o}-side region.

The next step consists in studying whether the asymptotic function 
also has a meaning if the normalization is not possible, e.g. in the Fisk-Stratonovich case, $\alpha = 1/2$.
Since the concept of infinite ergodicity has only recently been brought into physics \cite{bar1,bar2,bar3}, before studying the previous problem for the case of geometric Brownian motion, we study it for a simpler example from statistical mechanics that will allow us to better introduce the concept of infinite ergodicity. 

\section{Infinite ergodic theory in statistical mechanics}
\label{iet}

\begin{figure*}[t]
\centering
\includegraphics[scale=0.42]{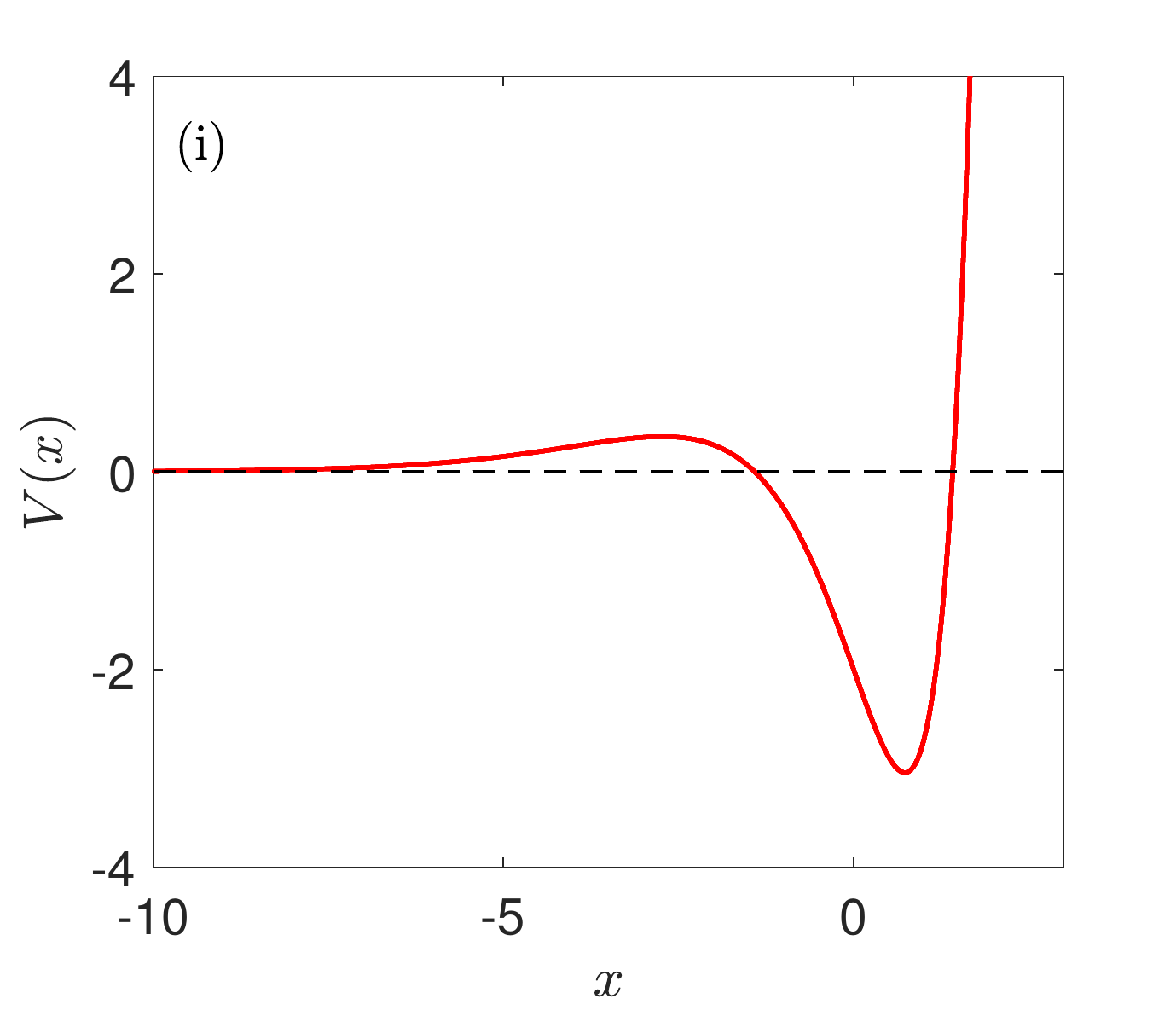}
\includegraphics[scale=0.42]{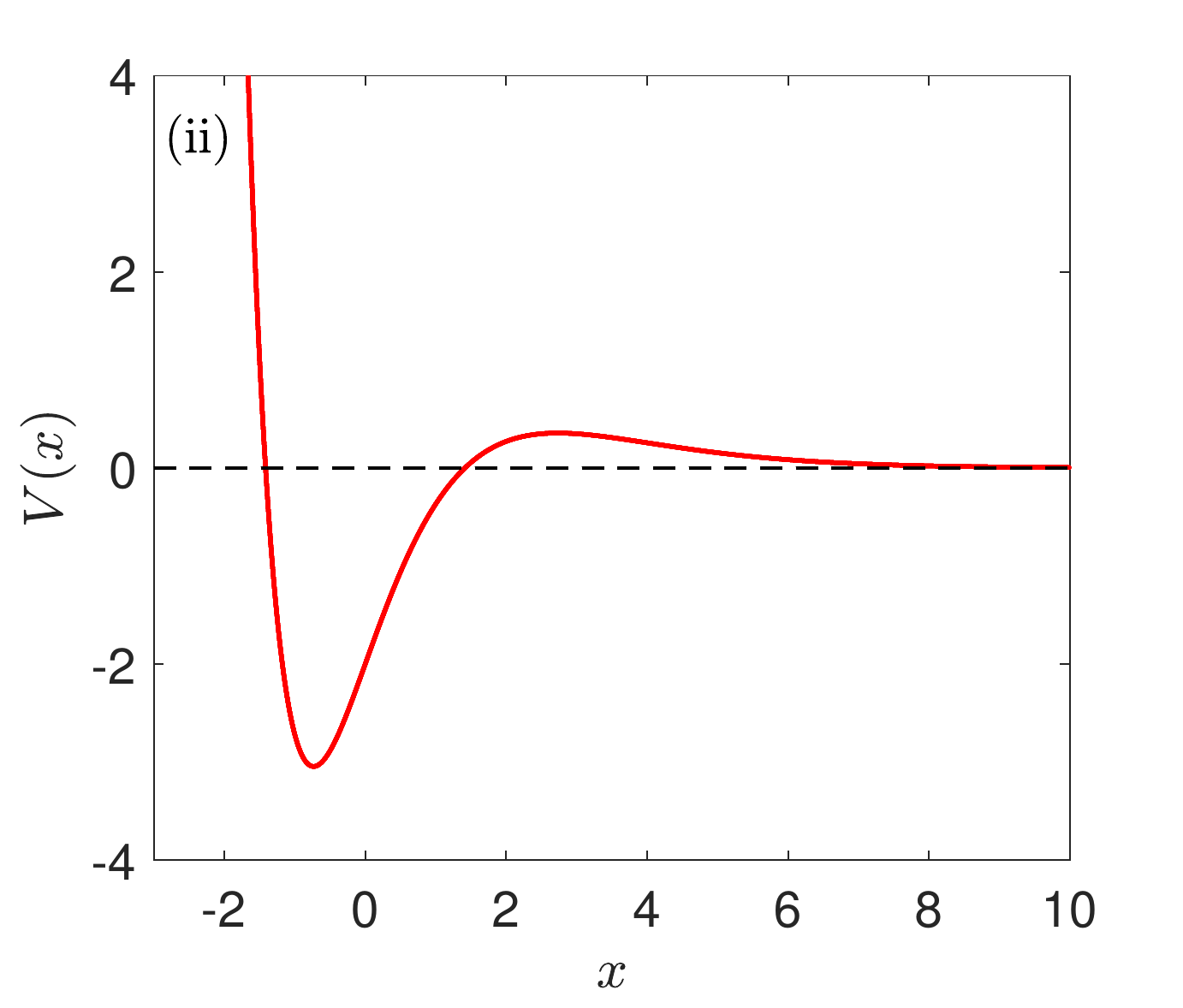}
\includegraphics[scale=0.42]{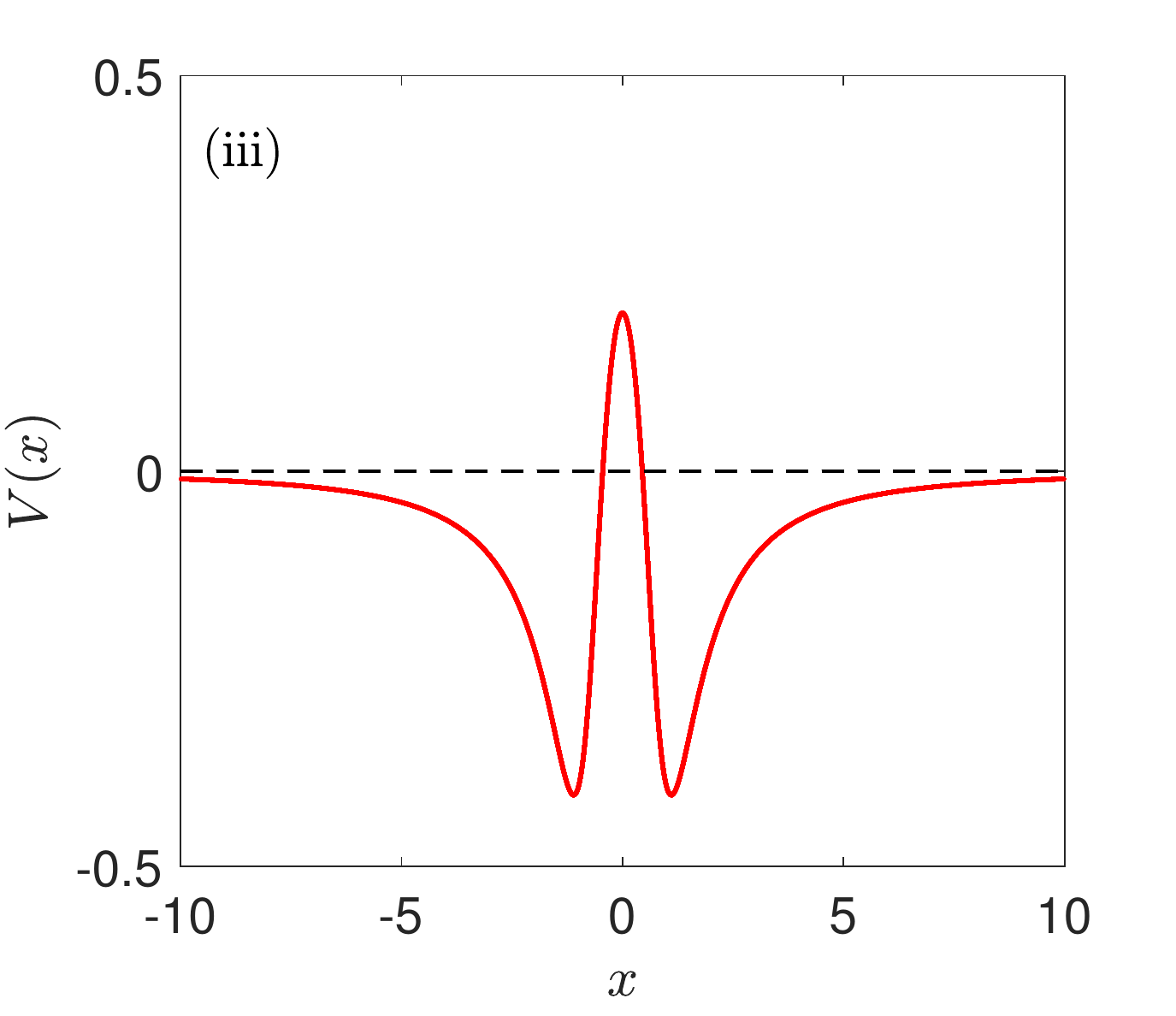}
\caption{\label{pots} Three examples of potential energy involving infinite ergodicity: (i) non-confining on the left, $\lim_{x\to -\infty}V(x)=0$ and $\lim_{x\to +\infty}V(x)=+\infty$; (ii) non-confining on the right, $\lim_{x\to -\infty}V(x)=+\infty$ and $\lim_{x\to +\infty}V(x)=0$; (iii) non-confining on both the left and the right, $\lim_{x\to -\infty}V(x)=0$ and $\lim_{x\to +\infty}V(x)=0$. }
\end{figure*}

In order to introduce the concept of infinite ergodicity into our discussion, in this Section we consider a simpler statistical mechanics system, following Refs.\cite{bar1,bar3}. 
Let us consider a particle of mass $m$ undergoing a one-dimensional overdamped stochastic motion under the effect of a potential energy $V(x)$. The Langevin equation reads as
\begin{equation}
\frac{\mathrm{d}x}{\mathrm{d}t}=-\frac{1}{m\gamma}\frac{\mathrm{d}V}{\mathrm{d}x}+\sqrt{\frac{k_BT}{m\gamma}}\xi(t),
\end{equation}
where $\gamma$ is the friction coefficient for unit mass, $k_B$ is the Boltzmann constant, $T$ is the temperature, and  $\xi(t)$ is the noise with the same properties as given 
above; note, that in this case the noise is simply additive and not multiplicative. The Fokker-Planck or Smoluchowski equation for the density $W(x,t)$ is given by \cite{man,gio}
\begin{equation}
\frac{\partial W}{\partial t}=\frac{\partial}{\partial x}\left( \frac{1}{m\gamma}\frac{\mathrm{d}V}{\mathrm{d}x}W+\frac{k_BT}{m\gamma}\frac{\partial W}{\partial x}\right).
\label{smol}
\end{equation}
As before, we search for the equilibrium distribution $W_{as}(x)$ through the equation
\begin{equation}
0= \frac{1}{m\gamma}\frac{\mathrm{d}V}{\mathrm{d}x}W_{as}+\frac{k_BT}{m\gamma}\frac{\mathrm{d} W_{as}}{\mathrm{d} x},
\end{equation}
which is solved by the Boltzmann distribution
\begin{equation}
W_{as}(x)=Ke^{-\frac{1}{k_BT}V(x)},
\label{boltz}
\end{equation}
where $1/K$ is the classical partition function. This density makes sense only if the partition function $\int \exp\left( -\frac{1}{k_BT}V\right) \mathrm{d}x$ 
converges in the region of interest. Typically, problems of non-convergence often emerge when non-confining potentials are used.
We try here to give a physical meaning to the obtained asymptotic density even when it is not normalizable. 
If $V(x)=0 $, of course, the partition function is not convergent but the general solution of the Fokker-Planck equation (which then reduces to the heat equation) 
is known since it corresponds to a Ornstein-Uhlenbeck process \cite{uh1,uh2,lax}. The result is
\begin{equation}
W(x,t)=\sqrt{\frac{m\gamma}{4 \pi k_B T  t}}\exp\left[ -\frac{m\gamma}{4k_B T t}(x-\mu_0)^2\right] ,
\label{heat}
\end{equation}
with the initial density $W(x,0)=\delta(x-\mu_0)$.
We can now imagine that for long times, in a case with a non-convergent partition function, the PDF evolution is given by a combination 
of Eqs.(\ref{boltz}) and (\ref{heat})
\begin{equation}
W(x,t)\underset{t\to\infty}\sim\sqrt{\frac{m\gamma}{4\pi k_B T  t}}e^{-\frac{1}{k_BT}V(x)}e^{ -\frac{m\gamma}{4k_B T t}(x-\mu_0)^2},
\label{oltzheat}
\end{equation}
or rather
\begin{equation}
W(x,t)\underset{t\to\infty}\sim\sqrt{\frac{m\gamma}{4\pi k_B T  t}}e^{-\frac{1}{k_BT}V(x)},
\label{oltzheat1}
\end{equation}
where we have used the property $\lim_{t\to\infty}e^{ -\frac{m\gamma}{4k_B T t}(x-\mu_0)^2} =1$.
Let us also consider that $V(x)\to 0$ for $x\to +\infty$ and/or $x\to -\infty$ in correspondence with the non-confining regions of the potential energy. 
In these regions we have a diffusive behavior of the system since the drift is negligible. 
The explored phase space is therefore infinite.
In order to verify the conjecture in Eq.(\ref{oltzheat1})
we have to demonstrate that the same expression is the solution of the Fokker-Planck equation for long times. From Eq.(\ref{oltzheat1}) the left hand side of Eq.(\ref{smol}) is obtained as
 \begin{equation}
\frac{\partial W(x,t)}{\partial t}\underset{t\to\infty}\sim - \sqrt{\frac{m\gamma}{16\pi k_B T  t^3}}e^{-\frac{1}{k_BT}V(x)}.
\label{oltzheat1der}
\end{equation}
Moreover, it is verified that the right hand side of Eq.(\ref{smol}) is exactly zero when calculated with Eq.(\ref{oltzheat1}). 
This indeed proves what is sought, since the term $\frac{\partial W(x,t)}{\partial t}$ goes to zero as $t^{-3/2}$, which is 
much faster than $t^{-1/2}$ (the leading term when $t\to\infty$), and is therefore negligible for long times, where  we search for the solution.
The remarkable point is that, from Eq.(\ref{oltzheat1}), we can write
\begin{equation}
\lim_{t\to\infty}\sqrt{\frac{4 \pi k_B T  t}{m\gamma}}W(x,t)= e^{-\frac{1}{k_BT}V(x)},
\label{oltzheat2}
\end{equation}
a result giving an important role to the Boltzmann exponential also for the case with divergent partition function. 

Furthermore, we can define an observable $\mathcal{O}(x)$ and introduce its ensemble average as
\begin{equation}
\left\langle \mathcal{O}(x) \right\rangle (t)=\int_{-\infty}^{+\infty}\mathcal{O}(x)W(x,t)dx.
\label{ave}
\end{equation}
From Eq.(\ref{oltzheat2}), we can write
\begin{equation}
\lim_{t\to\infty}\sqrt{\frac{4 \pi k_B T  t}{m\gamma}}\left\langle \mathcal{O}(x) \right\rangle (t)=\int_{-\infty}^{+\infty}\mathcal{O}(x)e^{-\frac{1}{k_BT}V(x)}dx,
\label{aveinf}
\end{equation}
which represents the infinite ergodicity property and, again, restores a role for the Boltzmann exponential factor also for the case with a divergent PDF. 
It means that the non-confining potential generates an infinite phase space (whence the term infinity ergodicity) explored by a drift-diffusion process, whose asymptotic properties are described by Eq.(\ref{aveinf}).

For completeness we discuss the convergence of the integral in Eq.(\ref{aveinf}) for some forms of potential energy.
To begin we suppose that $\mathcal{O}(x)=V(x)$ and we consider three cases:

(i) The potential energy is non-confining on the left: $\lim_{x\to -\infty}V(x)=0$ and $\lim_{x\to +\infty}V(x)=+\infty$. 
In this case the convergence of the integral in Eq.(\ref{aveinf}) is handled by $V(x)$ for $x\to -\infty$, and by $e^{-\frac{1}{k_BT}V(x)}$ for $x\to +\infty$, see Fig.\ref{pots} (i).

(ii) The potential energy is non-confining on the right: $\lim_{x\to -\infty}V(x)=+\infty$ and $\lim_{x\to +\infty}V(x)=0$. 
In this case the convergence of the integral in Eq.(\ref{aveinf}) is handled by $e^{-\frac{1}{k_BT}V(x)}$ for $x\to -\infty$, and by $V(x)$ for $x\to +\infty$, see Fig.\ref{pots} (ii).

(iii) The potential energy is non-confining on both the left and the right: $\lim_{x\to -\infty}V(x)=0$ and $\lim_{x\to +\infty}V(x)=0$. 
In this case the convergence of the integral in Eq.(\ref{aveinf}) is handled by  $V(x)$ for both $x\to -\infty$ and $x\to +\infty$, see Fig.\ref{pots} (iii).

The same discussion remains valid if we consider the force as observable, namely $\mathcal{O}(x)=-\mathrm{d}V(x)/\mathrm{d}x$. 
If, as an example, we consider a potential energy non-confining on the right (with $\lim_{x\to -\infty}V(x)=+\infty$ and $\lim_{x\to +\infty}V(x)=0$), 
we can write
\begin{eqnarray}
\nonumber
&&\lim_{t\to\infty}\sqrt{\frac{4 \pi k_B T  t}{m\gamma}}\left\langle -\frac{\mathrm{d}V}{\mathrm{d}x} \right\rangle =\int_{-\infty}^{+\infty}-\frac{\mathrm{d}V}{\mathrm{d}x}e^{-\frac{1}{k_BT}V}dx\\
\nonumber
&&=k_BT\int_{-\infty}^{+\infty}\frac{\mathrm{d}}{\mathrm{d}x}\left( e^{-\frac{1}{k_BT}V}\right) dx\\
&&=k_BT\left[ e^{-\frac{1}{k_BT}V(+\infty)}-e^{-\frac{1}{k_BT}V(-\infty)}\right]=k_B T,
\label{avef}
\end{eqnarray}
which is a constant, independent from the shape of the potential. If we divide Eq.(\ref{avef}) by the characteristic thermal length $\sqrt{K_BT/m}/\gamma$, we get
 \begin{eqnarray}
\lim_{t\to\infty}\sqrt{4 \pi \gamma t}\left\langle -\frac{\mathrm{d}V}{\mathrm{d}x} \right\rangle =\gamma\sqrt{k_B T m},
\label{avef1}
\end{eqnarray}
which has the physical units of force.
For further details concerning the infinite ergodic concept we refer to Refs.\cite{bar1,bar2,bar3}.

\section{Infinite ergodicity in geometric Brownian motion}
\label{ietgbm}

\begin{figure}[t]
\centering
\includegraphics[scale=0.6]{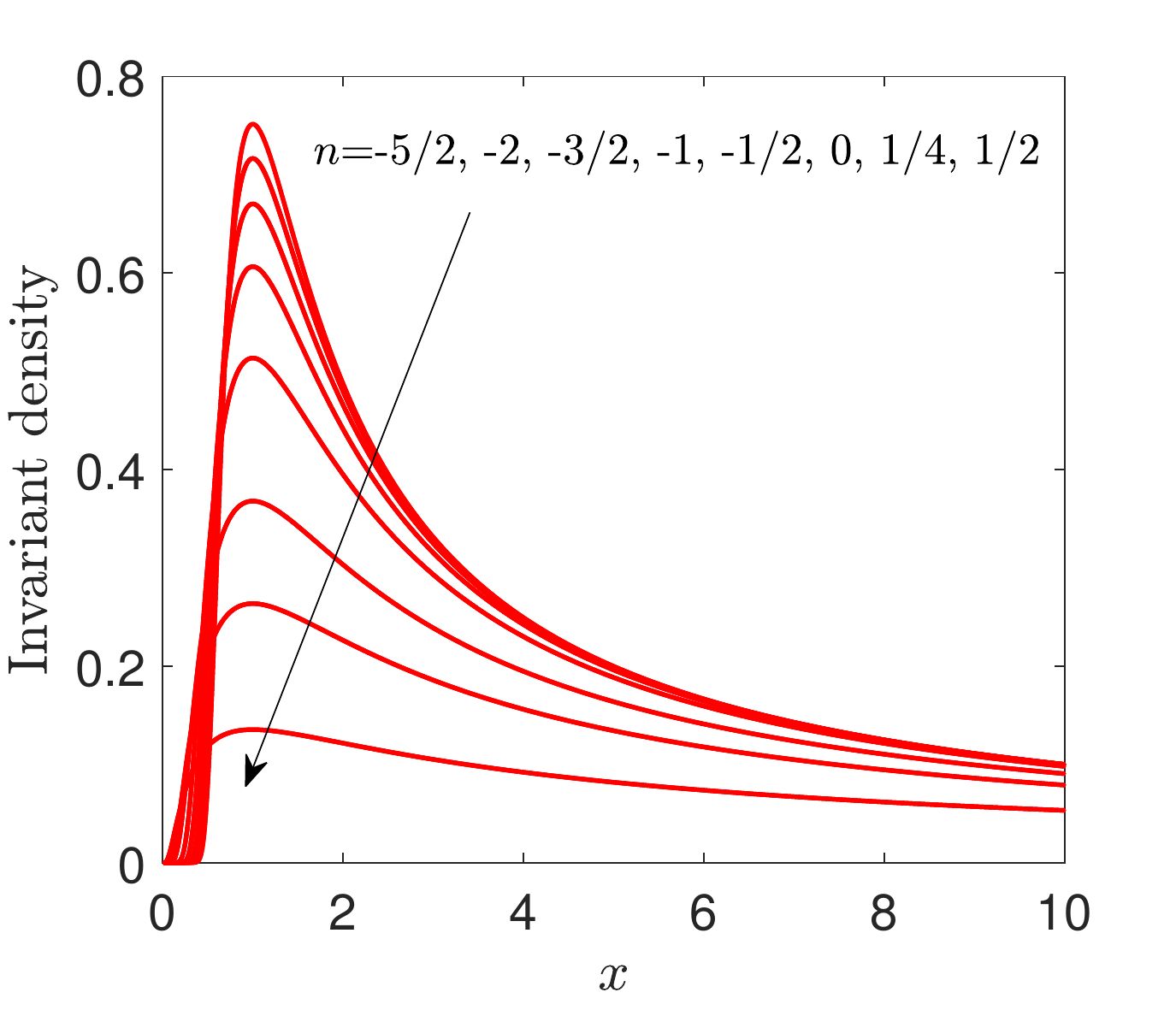}
\caption{\label{inva1} Invariant density defined in Eq.(\ref{asymp2}) for the geometric Brownian motion with nonlinear drift. We used the parameters $H_0$=-1, $G_0$=1 and a variable exponent $n$.  }
\end{figure}

By invoking the infinite ergodicity concept discussed in the previous Section, we now try to give significance to the non-normalized asymptotic solutions for the case $\alpha=1/2$ (Fisk-Stratonovich interpretation) in the equation 
\begin{equation}
\frac{\mathrm{d}x}{\mathrm{d}t}= - H_0x^n+G_0x\xi(t).
\label{tvgbmdriftpow}
\end{equation}
We still consider the relationship $G_0^2\frac{n-1}{H_0} > 0$ to be valid, through the hypotheses $ n - 1 < 0 $ and $ H_0 < 0 $. 
Hence, from Eq.(\ref{wwwa}), the non-normalized asymptotic density takes the form
\begin{equation}
W_{as}(x)\sim\frac{1}{x}\exp\left(-\frac{H_0}{G_0^2}\frac{x^{n-1}}{n-1} \right).
\label{wwwasy}
\end{equation}
We remark that the exponential term approaches 1 for $x\to\infty$ because of the assumptions $ n-1 < 0$ and $H_0 < 0$, exactly like the Boltzmann exponential of the previous Section in the non-confining regions.  
From the previously developed theory of geometric Brownian motion, we know that without drift we have the exact solution of the Fokker-Planck equation given by Eq.(\ref{solfp0}) (with $H_0=0$). In fact, with  $\alpha=1/2$ and $H_0=0$ we get
 \begin{equation}
W(x,t)=\frac{\exp\left[ {-\frac{\left(\log \frac{x}{\mu_0} \right)^2 }{4 G_0^2 t}}\right] }{2xG_0\sqrt{\pi t}},
\label{solfp0inf}
\end{equation}
corresponding to the initial condition $W(x,0)=\delta(x-\mu_0)$.
In analogy with the treatment of the overdamped Langevin equation in the previous section,
we can here assume a solution for long times of the process with $H_0\neq 0$ and $\alpha=1/2$ as a combination of Eqs.(\ref{wwwasy}) and (\ref{solfp0inf}). 
We therefore have for long times
\begin{equation}
W(x,t)\underset{t\to\infty}\sim\frac{\exp\left[ {-\frac{\left(\log \frac{x}{\mu_0} \right)^2 }{4 G_0^2 t}}\right] }{2xG_0\sqrt{\pi t}}\exp\left(-\frac{H_0}{G_0^2}\frac{x^{n-1}}{n-1} \right),
\label{asymp}
\end{equation}
or, equivalently,
\begin{equation}
W(x,t)\underset{t\to\infty}\sim\frac{1 }{2xG_0\sqrt{\pi t}}\exp\left(-\frac{H_0}{G_0^2}\frac{x^{n-1}}{n-1} \right),
\label{asymp1}
\end{equation}
where we have used the limiting property $\lim_{t\to\infty}\exp\left[ -\left(\log \frac{x}{\mu_0} \right)^2 /(4 G_0^2 t)\right]=1$.
To verify this conjecture we have to establish that Eq.(\ref{asymp1}) actually is the solution for long times of the following Fokker-Planck equation
\begin{equation}
\frac{\partial W}{\partial t}=H_0\frac{\partial }{\partial x}\left(x^nW \right)  + G_0^2\frac{\partial }{\partial x}\left[ x\frac{\partial }{\partial x}\left( xW\right)\right]   .
\label{FP3}
\end{equation}
By substituting Eq.(\ref{asymp1}) into Eq.(\ref{FP3}), we see that all the terms behaving as $t^{-1/2}$ (the leading terms) 
cancel each other out and only one negligible term remains of order $t^{-3/2}$. This term is again negligible as it tends to zero much faster than the others and 
therefore is not relevant for long times. Now, from Eq.(\ref{asymp1}) we obtain the important expression
\begin{equation}
\lim_{t\to\infty}2G_0\sqrt{\pi t}W(x,t)=\frac{1}{x}\exp\left(-\frac{H_0}{G_0^2}\frac{x^{n-1}}{n-1} \right).
\label{asymp2}
\end{equation}
Here, the right hand side is the so-called invariant density (see Fig.\ref{inva1}).
Also in this case we can define an arbitrary observable $\mathcal{O}(x)$ and introduce its expectation value (as an ensemble average) as
\begin{equation}
\left\langle \mathcal{O}(x) \right\rangle (t)=\int_{0}^{+\infty}\mathcal{O}(x)W(x,t)dx,
\label{avenew}
\end{equation}
where we considered the integration interval $(0,+\infty)$ to be consistent with the geometric Brownian motion phase space. 
Asymptotically, we get
\begin{equation}
\lim_{t\to\infty}2G_0\sqrt{\pi t}\left\langle \mathcal{O}(x) \right\rangle (t)=\int_{0}^{+\infty}\frac{\mathcal{O}(x)}{x}\exp\left(-\frac{H_0}{G_0^2}\frac{x^{n-1}}{n-1} \right)dx.
\label{avenew1}
\end{equation}

We now give an application of Eq.(\ref{avenew1}) with an observable defined as the power $\mathcal{O}(x)=x^s$ (where $s$ is a real number).
For this we simply rewrite the infinite ergodicity expression as
\begin{equation}
\lim_{t\to\infty}2G_0\sqrt{\pi t}\left\langle x^s \right\rangle (t)=\int_{0}^{+\infty}x^{s-1}\exp\left(-\frac{H_0}{G_0^2}\frac{x^{n-1}}{n-1} \right)dx,
\label{avenews}
\end{equation}
where the integral converges if $s/(n-1)>0$, provided that $G_0^2(n-1)/H_0>0$. This is true since the integral has the same form discussed in Eq.(\ref{wwwak}).
So, we have the closed-form expression
 \begin{equation}
\lim_{t\to\infty}2G_0\sqrt{\pi t}\left\langle x^s \right\rangle (t)=\frac{1}{\vert n-1 \vert}\left(\frac{n-1}{H_0} G_0^2\right)^\frac{s}{n-1}\Gamma\left( \frac{s}{n-1}\right),
\label{avenewsclosed}
\end{equation}
coming from Eq.(\ref{wwwakkk}).
An interesting special case corresponds to $s=n-1$ and yields
\begin{equation}
\lim_{t\to\infty}2G_0\sqrt{\pi t}\left\langle x^{n-1} \right\rangle (t)=\frac{\mbox{sgn}(n-1)}{H_0} G_0^2=\frac{1}{\vert H_0 \vert} G_0^2,
\label{avenewsclosed1}
\end{equation}
where $\mbox{sgn}(z)$ is the signum function extracting the sign of the real number $z$, and where we considered that $n-1<0$ and $H_0<0$. 
We note that the result in Eq.(\ref{avenewsclosed1}) is independent of $n$, i.e. independent of the shape of the forcing term in Eq.(\ref{tvgbmdriftpow}).
This result can be put in analogy with the asymptotic property of the average value of the force obtained in Eq.(\ref{avef1}).
Indeed, if we rewrite the stochastic differential equation in Eq.(\ref{tvgbmdriftpow}) in the form
\begin{equation}
\frac{1}{x}\frac{\mathrm{d}x}{\mathrm{d}t}=-H_0x^{n-1}+G_0\xi(t),
\end{equation}
we can identify the observable $x^{n-1}$ exactly as the force acting on the system.

\section{A further generalization} 
\label{further}

We finally consider the generalization of the geometric Brownian motion given in Eq.(\ref{tvgbmdriftspecial}), where the multiplicative noise term is now 
given by a nonlinear power with exponent $m$, as stated in the Introduction in Eq.(\ref{alg}),
\begin{equation}
\frac{\mathrm{d}x}{\mathrm{d}t}=-H_0x^n+G_0x^m \xi(t).
\label{tvgbmdriftpower}
\end{equation}
This equation represents a generalization with a drift term of the equation considered in Ref.\cite{bar3}, which was, except for the notation, 
of the form $\mathrm{d}x/\mathrm{d}t = G_0 x^m \xi(t)$.
The stochastic problem in Eq.(\ref{tvgbmdriftpower}) is associated with the Fokker-Planck equation
\begin{equation}
\frac{\partial W}{\partial t}=H_0\frac{\partial }{\partial x}\left(x^nW \right)  + G_0^2\frac{\partial }{\partial x}\left\lbrace  x^{2m\alpha}\frac{\partial }{\partial x}\left[ x^{2m(1-\alpha)}W\right]\right\rbrace   .
\label{FP4}
\end{equation}
For now, we first only assume that $m\neq 1$ in order to not reconsider the case already studied and 
search for an asymptotic solution $W_{as}(x)$ for the Fokker-Planck equation
\begin{equation}
0=H_0 x^nW_{as} + G_0^2  x^{2m\alpha}\frac{\partial }{\partial x}\left[ x^{2m(1-\alpha)}W_{as}\right] .
\label{FP4as}
\end{equation}
With the same technique used to solve Eq.(\ref{asy}), we find
\begin{equation}
W_{as}(x)=\frac{K}{x^{2m(1-\alpha)}}\exp\left(-\frac{H_0}{G_0^2}\frac{x^{n-2m+1}}{n-2m+1} \right),
\label{solasy4}
\end{equation}
for which we have to require that $n-2m+1\neq 0$. As before, $W_{as}$ is normalizable with finite $K$ when the integral over $(0,+\infty)$ is finite:
\begin{equation}
\frac{1}{K}=\int_0^{+\infty}\frac{1}{x^{2m(1-\alpha)}}\exp\left(-\frac{H_0}{G_0^2}\frac{x^{n-2m+1}}{n-2m+1} \right)\mathrm{d}x.
\label{solasyinte}
\end{equation}
The analysis follows that performed for Eq.(\ref{wwwak}), and generalizes it for finite $m \neq 1$:

(i) the term $x^{-2m(1-\alpha)}$ is convergent for $x\to 0$ when $2m(1-\alpha)<1$. The exponential term ensures the convergence for $x\to\infty$ if $a>0$ and $n-2m+1>0$. 
Finally, Eq.(\ref{solasyinte}) is convergent if
\begin{equation}
2m(1-\alpha)<1, \,\,\,H_0>0, \mbox{ and } n-2m+1>0;
\label{c1}
\end{equation}

(ii) the term $x^{-2m(1-\alpha)}$ ensures the convergence for $x\to\infty$ if $2m(1-\alpha)>1$. 
The exponential term provides the convergence for $x\to 0$ if $a<0$ and $n-2m+1<0$. Hence,  Eq.(\ref{solasyinte}) is convergent also if 
\begin{equation}
2m(1-\alpha)>1, \,\,\,H_0<0, \mbox{ and } n-2m+1<0\, .
\label{c2}
\end{equation}

\begin{figure}[t]
\centering
\includegraphics[scale=0.6]{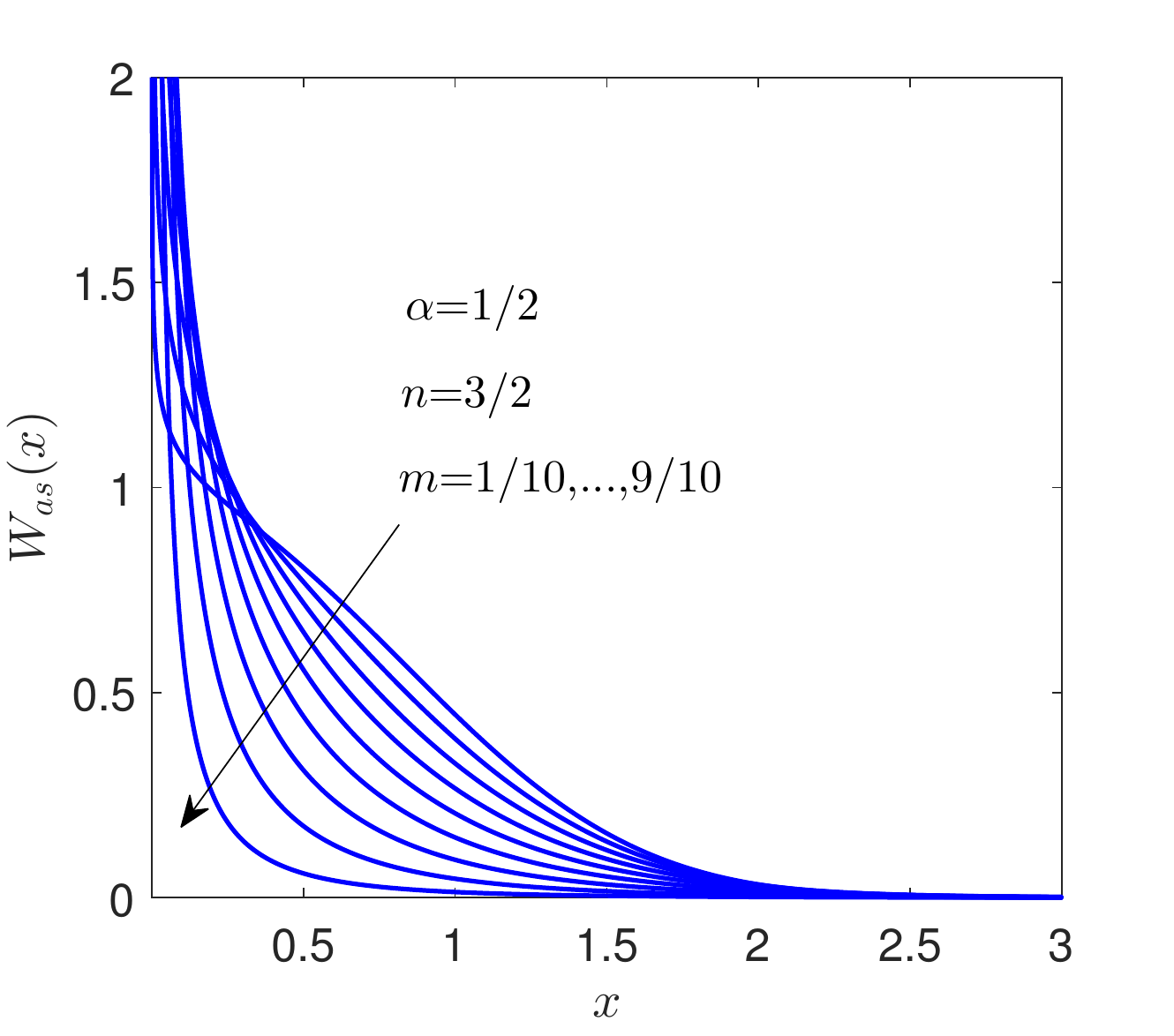}
\caption{\label{was2f} Example of asymptotic distributions in the  region defined by Eq.(\ref{c1}). We implemented Eq.(\ref{wwwafinpow}) with the parameters $H_0=3/2$ and $G_0=1$.}
\end{figure}

\begin{figure}[h]
\centering
\includegraphics[scale=0.6]{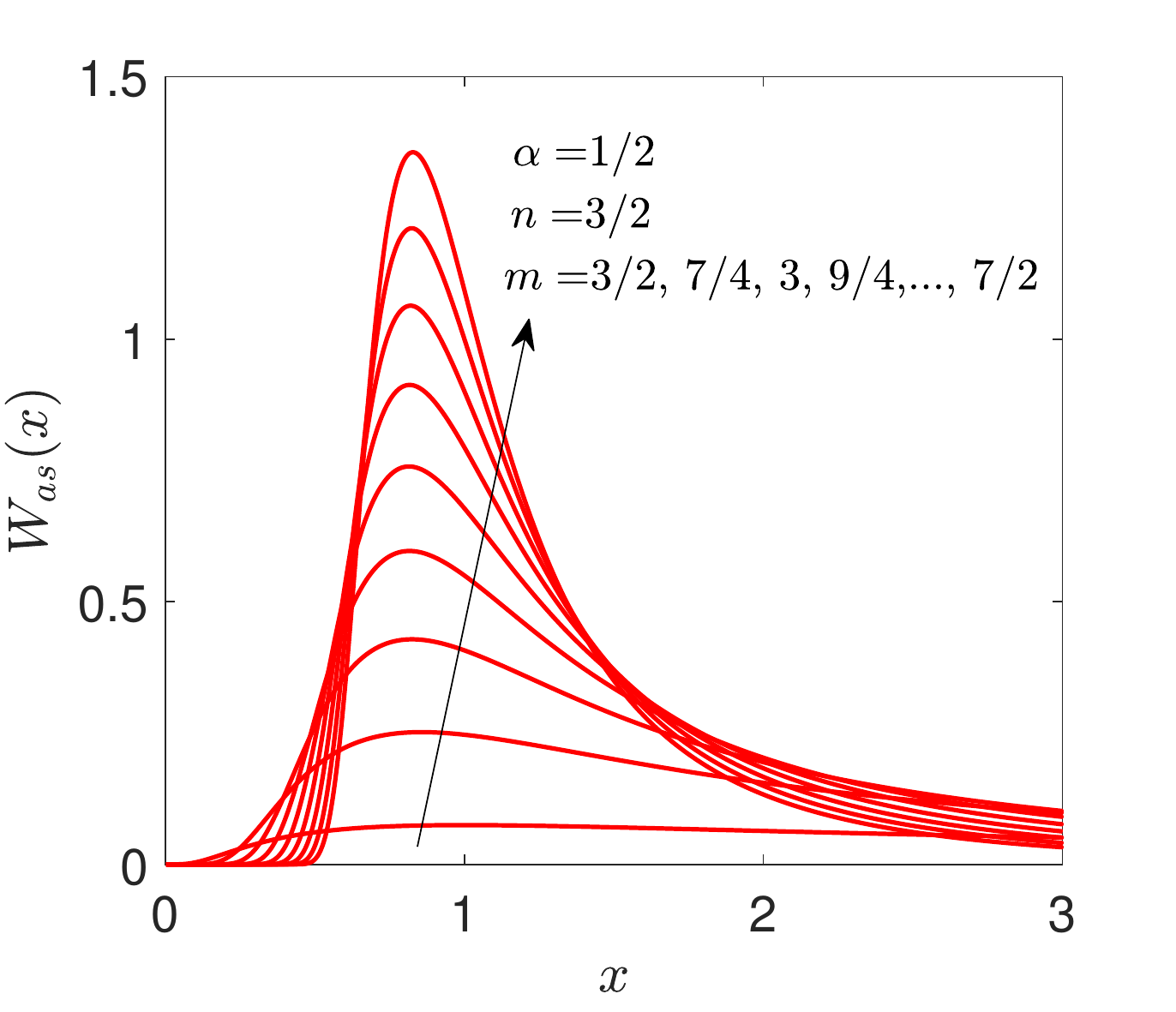}
\caption{\label{was1f} Example of asymptotic distributions in the It\^{o}-side region defined by Eq.(\ref{c2}). We implemented Eq.(\ref{wwwafinpow}) with the parameters $H_0=-3/2$ and $G_0=1$.}
\end{figure}

The calculation of the integral in Eq.(\ref{solasyinte}) can be done by the same method used before, and we get
\begin{equation}
\frac{1}{K}=\frac{\Gamma\left( \frac{2m\alpha-2m+1}{n-2m+1}\right)}{\vert n-2m+1 \vert}\left(\frac{n-2m+1}{H_0} G_0^2\right)^\frac{2m\alpha-2m+1}{n-2m+1} .
\label{wwwakkkpow}
\end{equation}
Therefore, the asymptotic density reads as
\begin{equation}
W_{as}(x)=\frac{\vert n-2m+1 \vert \exp\left(-\frac{H_0}{G_0^2}\frac{x^{n-2m+1}}{n-2m+1} \right)}{\left(\frac{n-2m+1}{H_0} G_0^2\right)^\frac{2m\alpha-2m+1}{n-2m+1}\Gamma\left( \frac{2m\alpha-2m+1}{n-2m+1}\right)x^{2m(1-\alpha)}},
\label{wwwafinpow}
\end{equation}
which is correct for 
\begin{equation}
G_0^2\frac{n-2m+1}{H_0}>0, \mbox{ and } \frac{2m\alpha-2m+1}{n-2m+1}>0,
\label{condspow}
\end{equation}
coming form Eqs.(\ref{c1}) and (\ref{c2}), and generalizing Eq.(\ref{conds}). In Figs.\ref{was2f} and \ref{was1f} we show the shape of the asymptotic density for different values of the parameters in the two regions, defined by Eqs.(\ref{c1}) and (\ref{c2}), respectively. We see that in the first region we have a singularity for $x=0$, whereas in the second region the density is regular everywhere. 

To obtain the infinite ergodic property for this system, we need to know the general solution of Eq.(\ref{tvgbmdriftpower}) without the forcing term, 
i.e. for $H_0=0$, for $m \neq 1$,
\begin{equation}
\frac{\mathrm{d}x}{\mathrm{d}t}=G_0x^m \xi(t).
\label{tvgbmdriftpowerspecial}
\end{equation}
This problem has been investigated in detail in Ref.\cite{bar2}, and the solution has been found by showing a connection with the 
so-called Bessel process \cite{bess1,bess2}. With our notation, we can say that the solution of Eq.(\ref{FP4}) with $H_0=0$, rewritten here as
\begin{equation}
\frac{\partial W}{\partial t}= G_0^2\frac{\partial }{\partial x}\left\lbrace  x^{2m\alpha}\frac{\partial }{\partial x}\left[ x^{2m(1-\alpha)}W\right]\right\rbrace ,
\label{FP5}
\end{equation}
is given by \cite{bar2}
\begin{eqnarray}
\label{solgene}
&&W(x,t)=\frac{\mu_0^{\frac{1}{2}(1-2m\alpha)}x^{\frac{1}{2}(1-4m+2m\alpha)}}{2G_0^2(1-m)t}\\
\nonumber
&&\times \exp\left[-\frac{\mu_0^{2(1-m)}+x^{2(1-m)}}{4 G_0^2(1-m)^2t} \right]I_{\frac{1-2m\alpha}{2(m-1)}}\left(\frac{\mu_0^{1-m}x^{1-m}}{2 G_0^2(1-m)^2t} \right) ,
\end{eqnarray}
for $x \geq 0$ (with reflecting boundary condition at $x=0$), and for the initial condition $W(x,0)=\delta(x-\mu_0)$.
Here $I_\nu(z)$ is the modified Bessel function of the first kind (of order $\nu$ and argument $z$) \cite{math1,math2,math3}. 
Importantly, this solution is valid when $0\leq m<1$ and $2m\alpha-2m+1>0$, and represents a time evolution that does not have a stationary PDF.
For $\alpha=1/2$ (Fisk-Stratonovich interpretation), we can use the relation \cite{math1,math2,math3}
\begin{equation}
I_{-\frac{1}{2}}(z)=\sqrt{\frac{2}{\pi z}}\cosh(z),
\end{equation}
and obtain the particular solution
\begin{eqnarray}
\label{solstra}
W(x,t)&=&\frac{1}{G_0\sqrt{\pi t} x^m}\cosh\left(\frac{\mu_0^{1-m}x^{1-m}}{2 G_0^2(1-m)^2t} \right) \\
\nonumber
&&\times \exp\left[-\frac{\mu_0^{2(1-m)}+x^{2(1-m)}}{4 G_0^2(1-m)^2t} \right].
\end{eqnarray}
This expression can be rewritten as 
\begin{eqnarray}
\label{solstrabis}
W(x,t)&=&\frac{1}{2G_0\sqrt{\pi t} x^m}\exp\left[-\frac{\left( x^{1-m}-\mu_0^{1-m}\right) ^2}{4 G_0^2(1-m)^2t} \right] \\
\nonumber
&&+\frac{1}{2G_0\sqrt{\pi t} x^m}\exp\left[-\frac{\left( x^{1-m}+\mu_0^{1-m}\right) ^2}{4 G_0^2(1-m)^2t} \right].
\end{eqnarray}
From the point of view of the physical interpretation, this form shows the superposition of an incident density (the first line) generated by the initial condition at $x=\mu_0$, and a 
reflected density (the second line) generated by the reflecting boundary condition at $x=0$. Moreover, if $m=0$, incident and reflected densities are Gaussian functions, as to be 
expected with additive noise.  

As always, the Stratonovich interpretation is closer to the physical understanding that can be attributed to the evolution of a stochastic system. 
In this case, the other interpretations with $\alpha\neq 1/2$ are able to break the symmetry between incident and reflected densities, as mathematically 
described by the Bessel function in Eq.(\ref{solgene}). Both Eqs.(\ref{solgene}) and (\ref{solstra}) can be proved by direct substitution in Eq.(\ref{FP5}).
By means of these solutions, we can study the asymptotic behavior, for large values of $t$, of the equation $\mathrm{d}x/\mathrm{d}t = G_0x^m \xi(t)$. 
To do this, we use the property \cite{math1,math2,math3}
\begin{equation}
I_{\nu}(z) \underset{z \to 0}\sim \left( \frac{1}{2} z \right)^\nu \frac{1}{ \Gamma (\nu+1)},
\end{equation}
and we obtain from Eq.(\ref{solgene})
\begin{eqnarray}
\label{solgeneasy}
W(x,t)\underset{t \to \infty}\sim\frac{[2(1-m)]^{(1-2\alpha)\frac{m}{1-m}}}{(G_0^2t)^{\frac{2m\alpha-2m+1}{2(1-m)}}\Gamma\left[ \frac{2m\alpha-2m+1}{2(1-m)}\right] x^{2m(1-\alpha)}}.
\end{eqnarray}
In the particular case with $\alpha=1/2$, we use the Gamma function value $\Gamma(1/2)=\sqrt{\pi}$ \cite{math1,math2,math3}, and we obtain from Eq.(\ref{solstra}) or Eq.(\ref{solstrabis}) the simpler asymptotic behavior
\begin{eqnarray}
\label{solstraasy}
W(x,t)\underset{t \to \infty}\sim\frac{1}{G_0\sqrt{\pi t} x^m}.
\end{eqnarray}

\begin{figure}[t]
\centering
\includegraphics[scale=0.6]{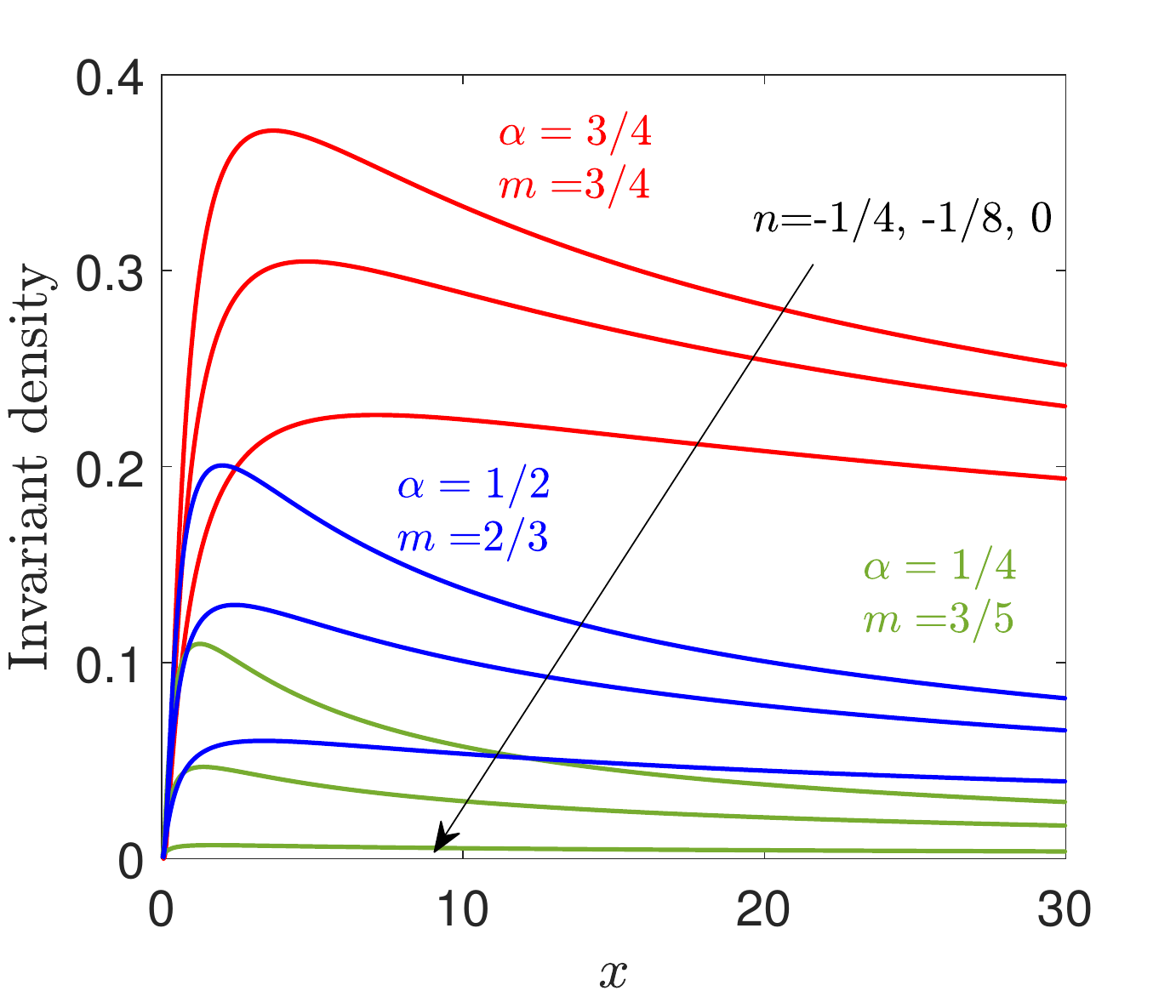}
\caption{\label{inva2} Invariant density defined in Eqs.(\ref{invm}) and (\ref{inv12})  for the generalized geometric Brownian motion with nonlinear drift. We used the parameters $H_0$=-1, $G_0$=1, variable exponents $n$ and $m$, and variable parameter $\alpha$.  }
\end{figure}

Summing up, on the one hand, we can say that the process with drift term, see Eq.(\ref{tvgbmdriftpower}), 
exhibits an equilibrium asymptotic solution when Eq.(\ref{c1}) or Eq.(\ref{c2}) is satisfied. On the other hand, for the equation without forcing term, 
see Eq.(\ref{tvgbmdriftpowerspecial}), there is no equilibrium and we know the asymptotic evolution when $0\leq m<1$ and $2m\alpha-2m+1>0$.
The idea of the infinite ergodicity is to give meaning to the equilibrium solution $W_{as}(x)$ of Eq.(\ref{tvgbmdriftpower}) even when it cannot be normalized. 
Hence, we consider the conditions $0\leq m<1$ and $2m\alpha-2m+1>0$, under which we know the asymptotic solution of Eq.(\ref{tvgbmdriftpowerspecial}), 
and we add the assumptions $H_0 < 0$ and $n-2m+1<0$, in such a way that $W_{as}(x)$ it is not normalizable. 
When this set of conditions is satisfied, we can try to merge Eqs.(\ref{wwwafinpow})  and (\ref{solgeneasy}) in order to get the asymptotic behavior. 
%of Eq.(\ref{tvgbmdriftpower}). 
This is facilitated by the fact that in both formulae there is the same power $x^{2m(1-\alpha)}$ in the denominator. We therefore propose to consider
\begin{eqnarray}
\label{solgeneasyfull}
W(x,t)\underset{t \to \infty}\sim\frac{[2(1-m)]^{(1-2\alpha)\frac{m}{1-m}}\exp\left(-\frac{a}{G_0^2}\frac{x^{n-2m+1}}{n-2m+1} \right)}{(G_0^2t)^{\frac{2m\alpha-2m+1}{2(1-m)}}\Gamma\left[ \frac{2m\alpha-2m+1}{2(1-m)}\right] x^{2m(1-\alpha)}}.
\end{eqnarray}
If $\alpha=1/2$, we can merge Eqs.(\ref{wwwafinpow}) and (\ref{solstraasy}) and have
\begin{eqnarray}
\label{solstraasyfull}
W(x,t)\underset{t \to \infty}\sim\frac{\exp\left(-\frac{H_0}{G_0^2}\frac{x^{n-2m+1}}{n-2m+1} \right)}{G_0\sqrt{\pi t} x^m}.
\end{eqnarray}
These proposals should represent the asymptotic behavior of Eq.(\ref{tvgbmdriftpower}) when $0\leq m<1$, $2m\alpha-2m+1>0$,  $H_0<0$, and $n-2m+1<0$. 
The verification by direct substitution into the Fokker-Planck equation proceeds as before.
So, Eq.(\ref{solgeneasyfull})  can be rewritten as
\begin{eqnarray}
\nonumber
&&\lim_{t\to\infty}\frac{\Gamma\left[ \frac{2m\alpha-2m+1}{2(1-m)}\right](G_0^2t)^{\frac{2m\alpha-2m+1}{2(1-m)}}W(x,t)}{[2(1-m)]^{(1-2\alpha)\frac{m}{1-m}}}
\\&&\,\,\,\,\,\,\,\,\,\,\,\,\,\,\,\,\,\,\,\,\,\,\,\,\,\,\,\,\,\,=\frac{\exp\left(-\frac{H_0}{G_0^2}\frac{x^{n-2m+1}}{n-2m+1} \right)}{ x^{2m(1-\alpha)}}, 
\label{invm}
\end{eqnarray}
and Eq.(\ref{solstraasyfull}) for $ \alpha = 1/2$ as
\begin{equation}
\lim_{t\to\infty}G_0\sqrt{\pi t}W(x,t)=\frac{\exp\left(-\frac{H_0}{G_0^2}\frac{x^{n-2m+1}}{n-2m+1} \right)}{ x^m}.
\label{inv12}
\end{equation}
These results represent the asymptotic evolution of the density $W(x,t)$, solving Eq.(\ref{FP4}) for large values of $t$, in the case where there is no  equilibrium 
PDF, i.e. they represent the generalization of Eq.(\ref{asymp2}), obtained previously for the case of geometric Brownian motion. In particular, Eq.(\ref{inv12}) reduces to Eq.(\ref{asymp2}) when $m=1$. The right hand sides of Eqs.(\ref{invm}) and (\ref{inv12}) are therefore the so-called invariant densities of the system (see Fig.\ref{inva2}).

In an analogous fashion we can introduce an observable $\mathcal{O}(x)$ with its ensemble average defined in Eq.(\ref{avenew}). 
From previous asymptotic results, we easily obtain
\begin{eqnarray}
\nonumber
&&\lim_{t\to\infty}\frac{\Gamma\left[ \frac{2m\alpha-2m+1}{2(1-m)}\right](G_0^2t)^{\frac{2m\alpha-2m+1}{2(1-m)}}}{[2(1-m)]^{(1-2\alpha)\frac{m}{1-m}}}\left\langle \mathcal{O}(x) \right\rangle (t) \\
&&\,\,\,\,\,\,\,\,\,\,\,\,\,\,\,=\int_0^{+\infty}\frac{\exp\left(-\frac{H_0}{G_0^2}\frac{x^{n-2m+1}}{n-2m+1} \right)}{ x^{2m(1-\alpha)}}\mathcal{O}(x)\mathrm{d}x, 
\label{fin1}
\end{eqnarray}
for arbitrary values of $\alpha$, and
\begin{eqnarray}
\nonumber
\lim_{t\to\infty}G_0\sqrt{\pi t}\left\langle \mathcal{O}(x) \right\rangle (t)=\int_0^{+\infty}\frac{\exp\left(-\frac{H_0}{G_0^2}\frac{x^{n-2m+1}}{n-2m+1} \right)}{ x^m}\mathcal{O}(x)\mathrm{d}x,\\
\label{fin2}
\end{eqnarray}
for $\alpha=1/2$.
Both results are valid when $0\leq m<1$, $2m\alpha-2m+1>0$,  $H_0<0$, and $n-2m+1<0$.
These two results again give important significance to asymptotic distributions even when the latter cannot be normalized and therefore represent a further example of infinite ergodic theory. 
They are valid only when the form of the observable $\mathcal{O}(x)$ renders the integral appearing in Eqs.(\ref{fin1}) and (\ref{fin2}) convergent. 

\section{Conclusions}

We have studied the stochastic process of geometric Brownian motion and some of its generalizations as given by
Eq.(\ref{tvgbmdriftpower})
for general values of the drift exponent $n$, the diffusion exponent $m$, and the discretization parameter $ 0 \leq \alpha \leq 1$. The corresponding Fokker-Planck equation
is readily written down, following established procedures. The study of the asymptotic probability distributions of the Fokker-Planck equation
reveals that the normalizability of the PDF at large times is tied to general conditions on the exponents $m, n,$ and $\alpha$. 
We establish the conditions on the exponents $n,m$ and on the discretization parameter $\alpha$ for which this is the case.
Our - surprising - main finding for the case of the standard geometric Brownian noise with $m=1$ is that the presence of a 
drift term in the stochastic equation allows to produce normalizable stationary PDFs provided $\alpha \neq 1/2$. If $\alpha=1/2$ (Fisk-Stratonovich case), 
the concept of infinite ergodicity allows to derive a well-defined invariant density, defined on the right hand side of Eq.(\ref{asymp2}). In the generalizations for
$ m \neq 1$, our results link to the findings by Barkai and collaborators, notably those of Ref.\cite{bar2}. 
In this case, we are able to find an invariant density for $\alpha=1/2$ (Fisk-Stratonovich case), see the right hand side of Eq.(\ref{inv12}), but also another invariant density for an arbitrary stochastic interpretation, see Eq.(\ref{invm}). 
In conclusion, we can say that infinite ergodic theory provides interesting results not only for classical statistical mechanics with additive noise, but also for more complex stochastic processes with multiplicative noise such as geometric Brownian motion or its generalizations. More specifically, the obtained results allows us to exactly determine the asymptotic behavior of physical observables in complex drift-diffusion driven systems even though we cannot find the general solution of the associated Fokker-Planck equation.

\begin{acknowledgments}

S.G. likes to acknowledge discussions with Tom Dupont concerning the time-varying geometric Brownian motion. R.B. thanks 
Rainer Grauer for discussions on the use of geometric Brownian motion in the statistical theory of turbulence. All authors acknowledge
support funding of the French National Research Agency ANR through project `Dyprosome' (ANR-21-CE45-0032-02). 
\end{acknowledgments}
 
%\nocite{*}
%\bibliography{aipsamp}% Produces the bibliography \right) via BibTeX.

\end{document}